\documentclass[twocolumn,11pt]{article} 
\usepackage{amssymb}
\usepackage{amsmath}
\usepackage{epic}
\usepackage{eepic}
\usepackage{graphicx}
\DeclareSymbolFontAlphabet{\mathbb}{AMSb}
\usepackage{Vmargin}
\setmargins{1in}{1in}{6.5in}{9in}{0pt}{0pt}{12pt}{42pt}

\newcommand{\sqrtsz}{{\sigma_z}^{\frac{1}{2}}}
\newcommand{\sqrtsx}{{\sigma_x}^{\frac{1}{2}}}

\begin{document} 

\title{On Universal and Fault-Tolerant Quantum Computing
\thanks{This work 
was supported in part by grants from the Revolutionary Computing
group at JPL (contract \#961360), and from the DARPA Ultra program
(subcontract from Purdue University \#530--1415--01).}}

\author{P. Oscar Boykin, Tal Mor, Matthew Pulver,
Vwani Roychowdhury, and Farrokh Vatan
\thanks{E--mail addresses of the authors are, respectively:
\tt \{boykin, talmo, pulver, vwani, vatan\}@ee.ucla.edu.} 
\\ \small Electrical Engineering Department \\ \small UCLA
\\ \small Los Angeles, CA 90095}

\date{}

\maketitle

\begin{abstract}
A novel universal and fault-tolerant basis (set of gates)
for quantum computation is described.  Such a set is necessary to
perform quantum computation in a realistic noisy environment. The new
basis consists
of two single-qubit gates (\emph{Hadamard} and
${\sigma_z}^{\frac{1}{4}}$), and one double-qubit gate (Controlled-NOT).
Since the set consisting of Controlled-NOT and Hadamard gates is not
universal, the new basis achieves universality by including only
one additional elementary (in the sense that it does not include
angles that are irrational multiples of $\pi$) single-qubit gate,
and hence, is 
potentially the simplest universal basis that one can construct.
We also provide an alternative
proof of universality for the {\em only other known class} of
universal and fault-tolerant basis proposed in \cite{shorFT,kitaevQC}.
\end{abstract}

\section{Introduction} 

A new model of computation based on the laws of quantum mechanics
has been shown to be superior to standard (classical) computation
models~\cite{shorFactor, Grover}.  Potential realizations
of such computing devices are currently under extensive research
\cite{cirac:zoller,monroe,turchette,haroche,Cory,CG,NMR,ESR}, and the
theory of using them in a realistic noisy environment is still developing. 
Two of the main requirements for error-free operations
are to have a set of gates that is both universal for quantum computing
(see~\cite{egfqc} and references therein), and that can operate in
a noisy environment 
(i.e., fault-tolerant)~\cite{shorQEC,shorFT,aharonov1,KLZ,kitaevQC}.

A scheme to correct errors in quantum bits (qubits) was proposed
by Shor~\cite{shorQEC} by adopting standard coding techniques and
modifying them to correct quantum mechanical errors induced by the
environment.  In such quantum error-correction techniques, the two states of
each qubit are encoded using a string of qubits, so that the state of the
qubit is kept in a pre-specified two-dimensional subspace of the space spanned by the
string of qubits.  We refer to this as the logical qubit.  This is done
in a way that error in one or more (as permitted by the code) physical 
qubits will not destroy the logical qubit.
To avoid errors in the computation itself, 
Shor~\cite{shorFT} suggested performing  the
computations on the logical qubits (without first decoding them), and this
type of computation is known as fault-tolerant computation. 

There are a number of requirements that a fault-tolerant quantum circuit
must satisfy. To prevent propagation of single-qubit errors to other
qubits in the same code word, one requirement of fault-tolerant computation is to 
disallow operations between any two qubits from the same codeword. This constraint 
imposes significant restrictions on both the types of unitary operations that can be performed on the
encoded logical qubits, and the quantum error-correcting codes that can be used to
encode the logical qubits. For example, if a ``double-even'' CSS code\footnote{Double 
even codes are punctured  binary codes in which the number of ones in any
codeword is a multiple of 4 \cite{shorFT}.} (e.g.,
the $((7,1,3))$ quantum code described in \cite{calderbank,steane}) is
used then one can show that the following unitary operations can be fault-tolerantly implemented:
\begin{equation} H, \ \sigma_z^\frac{1}{2}, \
\wedge_1(\sigma_x) =\left(\begin{array}{rrrr}1&0&0&0\\0&1&0&0\\
0&0&0&1\\0&0&1&0\end{array}\right) \label{cliff-group} \end{equation}
where $H$ and $\sigma_z^\frac{1}{2}$ are defined in the next section,
and $\wedge_k(U)$ denotes the controlled--$U$ operation with $k$ control bits
(see \cite{egfqc}). So far, these operations are the only
ones that have been shown to be "directly" fault-tolerant (in the sense that
no measurements and/or preparations of special states are required) operations. 
It is well known, however, that the group generated by the above 
operations (also referred to as the {\em normalizer group}) is not universal for 
quantum computation. This leads to the interesting problem of determining a
basis that is both universal and can be implemented fault-tolerantly.

There are several well-established results on the universality of quantum bases
\cite{adleman, barenco, egfqc, bernstein, deutsch89}.  Proofs of universality
of these bases rest primarily on the fact that they include at least one
``non-elementary'' gate, i.e., a gate that performs a rotation on single qubits 
by an irrational multiple of $\pi$.  A direct fault-tolerant realization of such 
a gate, however, is not possible; this property makes all the well-known 
universal bases inappropriate for practical and noisy quantum computation.

The search for universal and fault-tolerant bases has led to a novel
basis as proposed in the seminal work of Shor \cite{shorFT}. It includes 
the Toffoli gate in addition to the above-mentioned generators of the 
normalizer group; hence, the basis can be represented by the following set
 $\{\,H,\, {\sigma_z}^{\frac{1}{2}},\,
\Lambda_2(\sigma_x)\,\}$. A fault-tolerant realization of the Toffoli gate 
(involving only the generators of the normalizer group, 
preparation of a special state, and appropriate measurements) has been shown
in \cite{shorFT}; a proof of universality of this basis, however, was not included. 
Later Kitaev \cite{kitaevQC} proved the universality of a basis, comprising the set 
$\{\wedge_1(\sqrtsz), H\}$ (see Section 5.1), that is ``equivalent''
to Shor's basis, i.e., the gates in the  new basis can be {\em exactly} realized using gates
in Shor's basis and vice-versa. This result provides the first published proof
of the universality of a fault tolerant basis.

A number of other researchers have proposed fault-tolerant bases
that are equivalent to Shor's basis.  Knill, Laflamme, and
Zurek \cite{KLZ96} considered the bases
$\{\, {\sigma_z}^{\frac{1}{2}},\, \Lambda_1({\sigma_z}^{\frac{1}{2}}),\, 
\Lambda_1(\sigma_x)\,\}$  
and 
$\{\, H,\, {\sigma_z}^{\frac{1}{2}},\, \Lambda_1({\sigma_z}^{\frac{1}{2}}),\, 
\Lambda_1(\sigma_x)\,\}$. The universality of these bases follows from the
fact that gates in Shor's basis can be simulated by small simple
circuits over these new bases. Hence, while novel fault-tolerant realizations
of the relevant gates in these bases were proposed, no new proofs of universality
was provided. 
The same authors later \cite{KLZ} studied a model in which the following set of gates
$\{\,H,\, {\sigma_z}^{\frac{1}{2}},\, \Lambda_1(\sigma_x)\,\}$ and the prepared
state $\cos(\pi/8)|0\rangle_L+\sin(\pi/8)|1\rangle_L$ \footnote{$|0\rangle_L$ 
and $|1\rangle_L$ refer to the states of the logical/encoded qubits.} are made
available. Again, the universality of this model follows from the fact 
that it can realize the gate $\Lambda_1(H)$, and consequently the Toffoli gate. 
We also note that
Aharonov and Ben-Or \cite{aharonov1} considered quantum systems with basic 
units that have $p>2$ states (referred to as
qupits). They proposed a class of quantum codes,
called polynomial codes, for such systems consisting of qupits. They defined a basis for the
polynomial codes and proved that it is universal. However their proof makes explicit use of
qupits with more than two states, and hence does not directly apply to the case studied in this
paper, where all operations are done on qubits only.

In this paper, we prove the existence of a novel basis
for quantum computation that lends itself to an elegant proof 
(based solely on the geometry of real rotations in three dimensions) of universality 
and in which all the gates can be easily realized in a fault-tolerant manner. 
In fact, we show that the inclusion of only one additional
single-qubit operation in the set in (\ref{cliff-group}), namely,
\[ {\sigma_z}^{\frac{1}{4}} \equiv  \left( \begin{array}{cc} 1 & 0 \\ 0 &
e^{i\frac{\pi}{4}} \end{array} \right) \]  
leads to a universal and fault-tolerant basis for quantum computation. 
Note that $ \sigma_z^\frac{1}{2}$ is not required anymore.
Thus, our basis consists of the following three gates
\begin{equation}
H, \ \sigma_z^\frac{1}{4}, \wedge_1(\sigma_x)\ . \label{our-basis}
\end{equation}
Proving the universality of Shor's basis (see \cite{kitaevQC}) seems to be a more 
involved process than proving the universality of the set of gates we suggest 
here. Moreover, we outline a general method for fault tolerant realizations of a 
certain class of unitary operations; the fault-tolerant realizations of both the
${\sigma_z}^{\frac{1}{4}}$ and the Toffoli gates are shown to be special cases of this
general formulation.

The first part of this paper is devoted to the proof of universality,
followed by a discussion on the fault-tolerant realization of the
${\sigma_z}^{\frac{1}{4}}$ gate, and finally an alternate proof for the
universality of Shor's basis. We also show in Appendix \ref{equiappendix} 
that the new basis proposed in this paper is not equivalent to Shor's basis.

\section{Definitions and Identities}
Define the usual Pauli matrices $\sigma_i$, and Hadamard operator $H$.\\

\begin{tabular}{ll}
$\sigma_x\equiv
\left(\begin{array}{cc} 0 & 1 \\ 1 &  0 \end{array} \right)\, ;$ &
\hspace*{-1.2in} $\sigma_z\equiv
\left(\begin{array}{rr} 1 & 0 \\ 0 & -1 \end{array} \right)\, ;$ \\ \\
$\sigma_y\equiv i\sigma_x\sigma_z = \left(\begin{array}{cc} 0 & -i \\ 
i &  0 \end{array} \right)\, ;$ & \\ \\
$H\equiv\frac{1}{\sqrt{2}}(\sigma_x+\sigma_z)= \frac{1}{\sqrt{2}}\left(
\begin{array}{cc} 1 & 1 \\ 1 &  -1 
\end{array} \right)\, .$ & \\ \\
\end{tabular}

First, we review some properties of matrices in $\mbox{\bf SU}(2)$. Note
that all traceless and Hermitian 2$\times$2 unitary matrices can be
represented as follows:
\[ \hat{n} \cdot \vec{\sigma}\equiv n_x
\sigma_x + n_y \sigma_y +n_z \sigma_z \ , \]
where $\hat{n}$ is a normalized real three-dimensional vector. From
commutation relations one can show that $\displaystyle 
{(\hat{n} \cdot \vec{\sigma})}^2=I$, and using this fact,
exponentiation of these Pauli matrices can be easily expanded as follows:
\[ e^{i\phi\hat{n} \cdot \vec{\sigma}}=\cos\phi\
I + i\sin\phi(\hat{n} \cdot \vec{\sigma})
\]
where $I$ is the 2$\times$2 identity matrix.  It should be noted that for
every element, $U$ in $\mbox{\bf SU}(2)$ there exist $\phi_{U}$ and
${\hat{n}}_U$,
such that \cite{SU2rep}
\[ U=e^{i\phi_U{\hat{n}}_U \cdot \vec{\sigma}}\; . \]

Next, to motivate our proof, we note the connection between real rotations
in three dimensions (i.e., elements of $\mbox{\bf SO}(3)$) and the group we
are concerned
with, $\mbox{\bf SU}(2)$. Note that Euler decompositions provide a way to 
represent a general rotation by an angle $2\phi$ about an axis $\hat{n}$, 
$R_{\hat{n}}(2\phi)$, by a product of rotations about two orthogonal
axes. That is,
\begin{equation}\label{euler0}
R_{\hat{n}}(2\phi)=
R_{\hat{z}}(2\alpha)R_{\hat{y}}(2\beta)R_{\hat{z}}(2\gamma).
\end{equation} 
There is a local isomorphism between $\mbox{\bf SO}(3)$ and $\mbox{\bf SU}(2)$.
For the same parameters in (\ref{euler0}), the following equation is
also true: 
\begin{equation}\label{euler1} e^{i\phi\hat{n} \cdot
\vec{\sigma}}=e^{i\alpha\sigma_z} e^{i\beta\sigma_y} e^{i\gamma\sigma_z} \ .
\end{equation} 
Thus, just as any rotation can be thought of as three rotations about
two axes, any element of $\mbox{\bf SU}(2)$ can be thought of as a product
of three
matrices, specifically, powers of exponentials of Pauli matrices.  In the
following section we will show that using the operations in 
our basis~(\ref{our-basis}),
we can approximate any ``rotation'' about two specific orthogonal axes,
and then by Euler decomposition, we will show how all elements of $\mbox{\bf
SU}(2)$
can be approximated.

Lastly, we introduce a set of notations involving real
powers of the Pauli matrices that prove to be very useful.
For example, there is an obvious way to raise $\sigma_z$ to a real power:
\[ \sigma_z^\alpha=\left(
\begin{array}{cc} 1 & 0 \\ 0 & e^{i\pi\alpha} \end{array} \right) \ .
\]
These matrices form an interesting family
for quantum computation, and are generally used to
put a relative phase between $|0\rangle$ and $|1\rangle$. For example,
 $\sigma_z^{\frac{1}{2^{n}}}$, for integer values of $n$, are used in Shor's
Factorization algorithm \cite{shorFactor}.  Based on the definition of
$\sigma_z^{\alpha}$, we use  similarity transformation to make the
following identities:
\[ \sigma_x^\alpha=
H\sigma_z^{\alpha}H, \ \
\sigma_y^{\alpha}=\sigma_z^{\frac{1}{2}}\sigma_x^{\alpha}\sigma_z^{-
\frac{1}{2}}, \ \
H^\alpha\equiv\sigma_y^{\frac{1}{4}}\sigma_z^{\alpha}\sigma_y^{-
\frac{1}{4}} .
\]
Note that we can also equivalently write $\displaystyle
\sigma_j^\alpha=e^{i\frac{\pi\alpha}{2}}e^{-
i\frac{\pi\alpha}{2}\sigma_j}$.

\section{A Proof of Universality} 
The proof of universality of our basis will be broken down into two
steps. In the first step we show that $H$ and $\sigma_z^{\frac{1}{4}}$
form a dense set in $\mbox{\bf SU}(2)$; i.e. for any element of $\mbox{\bf
SU}(2)$ and
desired degree of precision, there exists a finite product of $H$
and $\sigma_z^\frac{1}{4}$ that approximates it to this desired
degree of precision.  Next we observe that for universal quantum
computation all that is needed is $\wedge_1(\sigma_x)$ and $\mbox{\bf SU}(2)$
\cite{egfqc}.

For proving density in $\mbox{\bf SU}(2)$ using our basis, we first show
that we can
construct elements in our basis which correspond to rotations by angles
that are irrational multiples of $\pi$ in $\mbox{\bf SO}(3)$ about two
orthogonal
axes.  Once we have these irrational rotations about two orthogonal axes,
then the density in $\mbox{\bf SU}(2)$ follows simply from the local isomorphism
between $\mbox{\bf SU}(2)$ and $\mbox{\bf SO}(3)$ discussed in the previous
section. Let
\begin{equation} \begin{align} e^{i\lambda\pi\hat{n}_1 \cdot
\vec{\sigma}}&\equiv\sigma_z^{-\frac{1}{4}}\sigma_x^{\frac{1}{4}} \;
\hbox{ and } 
\label{en1} \\ e^{i\lambda\pi\hat{n}_2 \cdot \vec{\sigma}}&\equiv
H^{-\frac{1}{2}} \label{en2}
\sigma_z^{-\frac{1}{4}}\sigma_x^{\frac{1}{4}}H^{\frac{1}{2}}. \end{align}
\end{equation} 
Given (\ref{en1},\ref{en2}) one can calculate $\vec{n}_1$,
$\vec{n}_2$, and $\lambda$: 
\begin{equation} \begin{align}
\cos\lambda\pi&=\cos^2\frac{\pi}{8}=
\frac{1}{2}(1+\frac{1}{\sqrt{2}})\label{lambda} \\
\vec{n}_1&=(\sqrt{2}\cot\frac{\pi}{8})\frac{\hat{z}-
\hat{x}}{\sqrt{2}}+\hat{y}\label{n1}\\
\vec{n}_2&=(\sqrt{2}\cot\frac{\pi}{8})\hat{y}-\frac{\hat{z}-
\hat{x}}{\sqrt{2}}\label{n2} \end{align} \ ,\end{equation} 
where $\hat{x}, \hat{y}$, and $\hat{z}$ are the unit vectors along the
respective axes.
One can easily verify that $\vec{n}_1$, $\vec{n}_2$ are orthogonal.
(These would need to be normalized when used in exponentiation.)

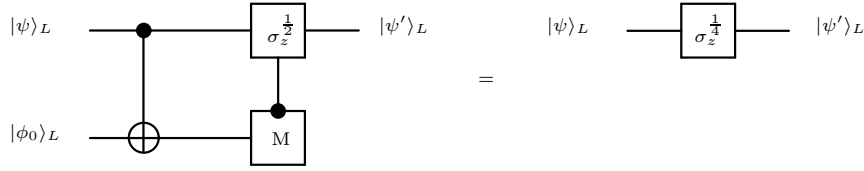
\begin{figure*}
\center
\setlength{\unitlength}{0.00083333in}
\begingroup\makeatletter\ifx\SetFigFont\undefined%
\gdef\SetFigFont#1#2#3#4#5{%
  \reset@font\fontsize{#1}{#2pt}%
  \fontfamily{#3}\fontseries{#4}\fontshape{#5}%
  \selectfont}%
\fi\endgroup%
\begin{picture}(5261,1045)(0,-10)
\thicklines
\put(838,851){\circle*{84}}
\put(838,180){\circle{188}}
\put(1677,348){\circle*{84}}
\drawline(1509,1018)(1844,1018)(1844,683)
	(1509,683)(1509,1018)
\drawline(1509,851)(503,851)
\drawline(838,851)(838,180)
\drawline(838,180)(838,96)
\drawline(503,180)(1509,180)
\drawline(1509,348)(1844,348)(1844,12)
	(1509,12)(1509,348)
\drawline(1677,683)(1677,348)
\drawline(1844,851)(2180,851)
\drawline(4192,1018)(4528,1018)(4528,682)
	(4192,682)(4192,1018)
\drawline(4528,850)(4863,850)
\drawline(3857,850)(4192,850)
\put(1635,138){\makebox(0,0)[lb]{\smash{{{\SetFigFont{7}{8.4}{\rmdefault}{\mddefault}{\updefault}M}}}}}
\put(2305,851){\makebox(0,0)[lb]{\smash{{{\SetFigFont{7}{8.4}{\rmdefault}{\mddefault}{\updefault}$|\psi'\rangle_L$}}}}}
\put(5030,851){\makebox(0,0)[lb]{\smash{{{\SetFigFont{7}{8.4}{\rmdefault}{\mddefault}{\updefault}$|\psi'\rangle_L$}}}}}
\put(3353,851){\makebox(0,0)[lb]{\smash{{{\SetFigFont{7}{8.4}{\rmdefault}{\mddefault}{\updefault}$|\psi\rangle_L$}}}}}
\put(2925,515){\makebox(0,0)[lb]{\smash{{{\SetFigFont{7}{8.4}{\rmdefault}{\mddefault}{\updefault}$=$}}}}}
\put(4276,780){\makebox(0,0)[lb]{\smash{{{\SetFigFont{7}{8.4}{\rmdefault}{\mddefault}{\updefault}$\sigma_z^{\frac{1}{4}}$}}}}}
\put(1610,780){\makebox(0,0)[lb]{\smash{{{\SetFigFont{7}{8.4}{\rmdefault}{\mddefault}{\updefault}$\sigma_z^{\frac{1}{2}}$}}}}}
\put(0,851){\makebox(0,0)[lb]{\smash{{{\SetFigFont{7}{8.4}{\rmdefault}{\mddefault}{\updefault}$|\psi\rangle_L$}}}}}
\put(0,180){\makebox(0,0)[lb]{\smash{{{\SetFigFont{7}{8.4}{\rmdefault}{\mddefault}{\updefault}$|\phi_0\rangle_L$}}}}}
\end{picture}
\caption{Implementing $\sigma_z^\frac{1}{4}$ fault-tolerantly.}
\label{sz4_meas}
\end{figure*}

The number $e^{i2\pi\lambda}$ is a root of the irreducible monic polynomial
\[ x^4+x^3+\frac{1}{4}x^2+x+1 \]
which is not cyclotomic (since not all coefficients are integers), 
and thus $\lambda$ is an irrational
number (see Appendix \ref{cycloappendix} Theorem \ref{cyclotheorem}).
Since $\lambda$ is irrational, it can be used to approximate any phase
factor $e^{i\phi}$:
\[ e^{i n\lambda\pi } \approx e^{i\phi} \]
for some $n\in\mathbb N$.

Now we at least have two dense subsets of $\mbox{\bf SU}(2)$. Namely,
$e^{i\alpha\hat{n}_1 \cdot \vec{\sigma}}$ and $e^{i\beta\hat{n}_2 \cdot
\vec{\sigma}}$: 
\begin{equation} \begin{split} \exists j \in \mathbb N \text{ such that }\lambda\pi j
\approx \alpha \pmod{2\pi} \\ \text{hence, } (e^{i\lambda\pi\hat{n}_1 \cdot
\vec{\sigma}})^j\approx e^{i\alpha\hat{n}_1 \cdot \vec{\sigma}}\ .
\end{split} \end{equation} 
\begin{equation} \begin{split} \exists j \in \mathbb N \text{ such that }
\lambda\pi k \approx \beta \pmod{2\pi}\\ \text{ hence, } (e^{i\lambda\pi\hat{n}_2 \cdot
\vec{\sigma}})^k\approx e^{i\beta\hat{n}_2 \cdot \vec{\sigma}} \ .
\end{split} 
\end{equation} 
Fortunately, this is all that is needed.
Since, $\vec{n}_1$ and $\vec{n}_2$ are orthogonal, we can write any
element of $\mbox{\bf SU}(2)$ in the following form \cite{SU2rep}:
\begin{equation}\label{euler} e^{i\phi\hat{n} \cdot
\vec{\sigma}}=(e^{i\alpha\hat{n}_1 \cdot \vec{\sigma}})
(e^{i\beta\hat{n}_2 \cdot \vec{\sigma}}) (e^{i\gamma\hat{n}_1 \cdot
\vec{\sigma}}) \ .\end{equation} 
Clearly, the representation in (\ref{euler}) is
analogous to Euler rotations about three orthogonal vectors. Expansion
of (\ref{euler}) gives: 
\begin{equation} \begin{align}
\cos\phi&=\cos\beta\cos(\gamma+\alpha)\label{cos}\\
\begin{split}\label{sn}
\hat{n}\sin\phi&=\hat{n}_1\cos\beta\sin(\gamma+\alpha)\\
&+\hat{n}_2\sin\beta\cos(\gamma-\alpha)\\
&+\hat{n}_1\times\hat{n}_2\sin\beta\sin(\gamma-\alpha) \end{split}
\end{align} 
\end{equation} 
For any element of $\mbox{\bf SU}(2)$ equations
(\ref{cos},\ref{sn}) can be inverted to find $\alpha$, $\beta$ and
$\gamma$. Since $\wedge_1(\sigma_x)$ and $\mbox{\bf SU}(2)$ form a universal
basis
for quantum computation \cite{egfqc}, it completes the proof
of universality of our basis.

There is no guarantee that, in general, it is possible to {\em efficiently 
approximate} an arbitrary phase $e^{i\phi}$ by repeated applications of the
available
phase $e^{i\pi \lambda}$. But using an argument similar to the one presented in
\cite{adleman}, we can show that for the given $\lambda$ (as defined in (\ref{lambda})), 
for any given $\varepsilon > 0$, with only
$\mbox{poly}(\frac{1}{\varepsilon})$ iterations of $e^{i\pi \lambda}$ we can
get $e^{i\phi}$ within $\varepsilon$. However, since our basis is already proven to be
universal, one can make use of an even better result.
As it is shown by Kitaev \cite{kitaevQC} and Solovay and Yao \cite{solovay},
every universal quantum basis $\cal B$ is efficient, 
in the sense that any unitary operation in $\mbox{U}(2^m)$, for constant $m$,
can be approximated within $\varepsilon$ by a circuit of size
$\mbox{poly-log}(\frac{1}{\varepsilon})$ over the basis $\cal B$.

\section{A Fault-Tolerant Realization of ${\sigma_z}^{\frac{1}{4}}$} 
We provide a simple scheme for the fault-tolerant realization 
of the ${\sigma_z}^{\frac{1}{4}}$ gate. The method describes a
general procedure that works for any quantum code for which the elements 
of the normalizer group can be implemented fault-tolerantly and involves 
the creation of special eigenstates of unitary transformations.

\begin{figure*}
\center
\setlength{\unitlength}{0.00083333in}
\begingroup\makeatletter\ifx\SetFigFont\undefined%
\gdef\SetFigFont#1#2#3#4#5{%
  \reset@font\fontsize{#1}{#2pt}%
  \fontfamily{#3}\fontseries{#4}\fontshape{#5}%
  \selectfont}%
\fi\endgroup%
\begin{picture}(3830,899)(0,-10)
\thicklines
\put(2000,812){\circle*{100}}
\drawline(1600,812)(1600,812)(1600,812)
	(1600,812)(1600,812)
\drawline(1200,812)(2800,812)
\drawline(1600,412)(1600,412)(1600,412)
	(1600,412)(1600,412)
\drawline(1600,412)(2400,412)(2400,12)
	(1600,12)(1600,412)
\drawline(1200,212)(1600,212)
\drawline(2400,212)(2800,212)
\drawline(2000,812)(2000,412)
\put(1740,162){\makebox(0,0)[lb]{\smash{{{\SetFigFont{8}{9.6}{\rmdefault}{\mddefault}{\updefault}$\sigma_z^{\frac{1}{4}}\sigma_x\sigma_z^{-\frac{1}{4}}$
}}}}}
\put(250,812){\makebox(0,0)[lb]{\smash{{{\SetFigFont{8}{9.6}{\rmdefault}{\mddefault}{\updefault}$|\vec{0}\rangle+|\vec{1}\rangle$}}}}}
\put(3000,462){\makebox(0,0)[lb]{\smash{{{\SetFigFont{8}{9.6}{\rmdefault}{\mddefault}{\updefault}$\alpha(|\vec{0}\rangle+|\vec{1}\rangle)|\phi_0\rangle_
L+\beta(|\vec{0}\rangle-|\vec{1}\rangle)|\phi_1\rangle_L$}}}}}
\put(0,212){\makebox(0,0)[lb]{\smash{{{\SetFigFont{8}{9.6}{\rmdefault}{\mddefault}{\updefault}$\alpha|\phi_0\rangle_L+\beta|\phi_1\rangle_L$}}}}}
\end{picture}
\caption{Creation of the $|\phi_0\rangle$eigenstate.}
\label{c_u_eigen}
\end{figure*}
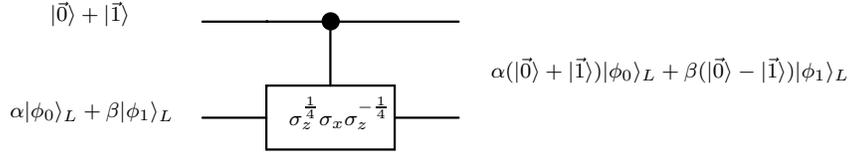

To perform $\sigma_z^{\frac{1}{4}}$ fault-tolerantly we use the following
state:
\begin{equation}\label{phi0}
|\phi_0\rangle\equiv\sigma_z^{\frac{1}{4}}H|0\rangle=\frac{|0\rangle +
e^{i\frac{\pi}{4}}|1\rangle}{\sqrt{2}} \ ,
\end{equation} 
(for which we later present the preparation process).
To apply $\sigma_z^{\frac{1}{4}}$ to a general single qubit state
$|\psi\rangle$ using this special state $|\phi_0\rangle$, first apply
$\wedge_1(\sigma_x)$ from $|\psi\rangle$ to $|\phi_0\rangle$.  See
Figure~\ref{sz4_meas}.
Then measure the second qubit ($|\phi_0\rangle$) in the computation basis.
If the result is $|1\rangle$, apply $\sigma_z^{\frac{1}{2}}$ 
to the
first qubit ($|\psi\rangle$). This leads to the desired operation,
as demonstrated in the following:
\begin{eqnarray*}
|\psi\rangle\otimes|\phi_0\rangle &=&
\mbox{$(\alpha|0\rangle+\beta|1\rangle)\otimes
 \frac{|0\rangle+e^{i\frac{\pi}{4}}|1\rangle}{\sqrt2} $}\\
 &\overset{\wedge_1(\sigma_x)}{\longrightarrow}&
  \mbox{$(\alpha|0\rangle+e^{i\frac{\pi}{4}}\beta|1\rangle)
   \otimes\frac{|0\rangle}{\sqrt{2}} $}\\
 &&\mbox{$+(\alpha|0\rangle+e^{-i\frac{\pi}{4}}\beta|1\rangle)
   \otimes e^{i\frac{\pi}{4}}\frac{|1\rangle}{\sqrt{2}} $}\\
 &\hspace*{-8mm} =& \hspace*{-8mm} \mbox{$\sigma_z^{\frac{1}{4}}|
 \psi\rangle\otimes\frac{|0\rangle}{\sqrt{2}}
  +\sigma_z^{-\frac{1}{4}}|\psi\rangle
   \otimes e^{i\frac{\pi}{4}}\frac{|1\rangle}{\sqrt{2}} $}
\end{eqnarray*}

Clearly, the above analysis shows that all that is necessary to perform
$\sigma_z^{\frac{1}{4}}$ fault tolerantly is the state $|\phi_0\rangle$
and the ability to do $\wedge_1(\sigma_x)$ and $\sigma_z^{\frac{1}{2}}$
fault-tolerantly. For CSS codes, $\wedge_1(\sigma_x)$, $H$ and
$\sigma_z^{\frac{1}{2}}$ can be done fault-tolerantly
\cite{shorFT,Gottesman-TFTC}. We next show how to generate the state 
$|\phi_0\rangle$ fault-tolerantly.

Fault tolerant creation of certain particular encoded eigenstates has been
discussed\cite{shorFT,KLZ}.
We present it in a more general way:
suppose that the fault-tolerant operation $U_\eta$ operates as follows:
\begin{equation}
U_\eta|\eta_i\rangle=(-1)^{i}|\eta_i\rangle
\end{equation}
on the states $|\eta_i\rangle$.
Thus, $U_\eta$ has the states $|\eta_i \rangle$ as 
eigenvectors with $\pm1$ as the eigenvalues.
Suppose we have access to a vector $|\psi\rangle$ such that:
\begin{equation}
|\psi\rangle=\alpha|\eta_0\rangle+\beta|\eta_1\rangle
\end{equation}
We show that using only bitwise operations, measurements, 
and this $|\psi\rangle$, the eigenvectors $|\eta_i\rangle$ can be obtained.
Now, to get the eigenvector of $U_\eta$ we make use of a $|\mbox{\sf
cat}\rangle$ state.
\begin{equation}
\mbox{$ |\mbox{\sf cat}\rangle=\frac{1}{\sqrt{2}}
(|00...0\rangle+|11...1\rangle)=\frac{1}{\sqrt{2}}(|\vec{0}\rangle+
|\vec{1}\rangle) $}
\end{equation}
See Figure~\ref{c_u_eigen}.
Applying $\wedge_1(U_\eta)$ bitwise, on $|\mbox{\sf
cat}\rangle\otimes|\psi\rangle$ we
obtain:
\begin{equation}
\mbox{$ |\mbox{\sf cat}\rangle\otimes|\psi\rangle\overset{\wedge_1(U_\eta)}
{\longrightarrow}\alpha(\frac{|\vec{0}\rangle+|\vec{1}\rangle}
{\sqrt{2}})|\eta_0\rangle+\beta(\frac{|\vec{0}\rangle-|\vec{1}\rangle}
{\sqrt{2}})|\eta_1\rangle $}
\end{equation}
A fault-tolerant measurement can be made to distinguish 
$\frac{|\vec{0}\rangle+|\vec{1}\rangle}{\sqrt{2}}$ 
from $\frac{|\vec{0}\rangle-|\vec{1}\rangle}{\sqrt{2}}$ \cite{shorFT}.
This measurement can be repeated to verify that you have it correct.

The fault-tolerant version of $\sigma_z^{\frac{1}{4}}$ needs 
the state $|\phi_0\rangle$, which can be generated using this formalism.  
In fact, $|\phi_0\rangle$ is an eigenstate of 
$U_\phi=\sigma_z^{\frac{1}{4}}\sigma_x\sigma_z^{-\frac{1}{4}}$.  
By commutation properties of 
 the $\sigma_z^{-\frac{1}{4}}$ operator, it is 
shown that $U_\phi$ can be realized with elements only from the normalizer Group.
\begin{equation}
U_\phi=\sigma_z^{\frac{1}{4}}\sigma_x\sigma_z^{-\frac{1}{4}}=
e^{i\frac{\pi}{4}}\sigma_z^{\frac{1}{2}}\sigma_x .
\end{equation}
Since $\sigma_z^{\frac{1}{2}}$ and $\sigma_x$ can be done fault-tolerantly, 
so can $U_\phi$.  We have claimed that $|\phi_0\rangle$ (\ref{phi0}) 
is an eigenvector, and now we state the other eigenvector.
\begin{equation}\label{phi1}
|\phi_1\rangle\equiv\sigma_z^{\frac{1}{4}}H|1\rangle=
\frac{|0\rangle - e^{i\frac{\pi}{4}}|1\rangle}{\sqrt{2}}
\end{equation}
One can verify now that these $|\phi_i\rangle$ are eigenvectors:
\begin{equation}
\begin{split}
U_\phi|\phi_i\rangle=\sigma_z^{\frac{1}{4}}
\sigma_x\sigma_z^{-\frac{1}{4}}|\phi_i\rangle \\ 
=\sigma_z^{\frac{1}{4}}\sigma_x\sigma_z^{-\frac{1}{4}}
(\sigma_z^{\frac{1}{4}}H|i\rangle)=\sigma_z^{\frac{1}{4}}
\sigma_x H|i\rangle \\ =\sigma_z^{\frac{1}{4}}H\sigma_z|i\rangle=
\sigma_z^{\frac{1}{4}}H(-1)^{i}|i\rangle=(-1)^{i}|\phi_i\rangle
\end{split}
\end{equation}
Since the $|\phi_i\rangle$ vectors are orthogonal, any single qubit state 
$|\psi\rangle$ can be represented as a sum of the $|\phi_i\rangle$.
\begin{equation}
|\psi\rangle=\alpha|0\rangle+\beta|1\rangle=
\alpha'|\phi_0\rangle+\beta'|\phi_1\rangle
\end{equation}
So, all the necessary ingredients are here: 
$|\psi\rangle$, and an appropriate fault-tolerant operation, $U_\phi$.  
If the outcome gives $|\phi_1\rangle$ rather than $|\phi_0\rangle$ 
we can flip the state:
\begin{equation}
\mbox{$ |\phi_0\rangle=\sigma_z|\phi_1\rangle=\sigma_z\frac{|0\rangle - 
e^{i\frac{\pi}{4}}|1\rangle}{\sqrt{2}}=\frac{|0\rangle + 
e^{i\frac{\pi}{4}}|1\rangle}{\sqrt{2}} $}
\end{equation}
Shor's implementation of Toffoli \cite{shorFT} also uses a special
case of this general procedure.  For performing Toffoli one uses
$U=\wedge_1(\sigma_z)\otimes\sigma_z$ to get the eigenstates:

\begin{eqnarray*}
|\mbox{\sf AND}\rangle &=&
\frac{1}{2}(|000\rangle+|010\rangle+|100\rangle+|111\rangle) \\
|\mbox{\sf NAND}\rangle &=&
\frac{1}{2}(|001\rangle+|011\rangle+|101\rangle+|110\rangle)
\end{eqnarray*}
Shor uses the $|\psi\rangle$ state of:
\begin{eqnarray*}
|\psi\rangle &=&
\frac{1}{\sqrt{2}}(|\mbox{\sf AND}\rangle+|\mbox{\sf NAND}\rangle) \\
             &=& (H|0\rangle)\otimes(H|0\rangle)\otimes(H|0\rangle)
\end{eqnarray*}
Thus the special state in \cite{shorFT} 
can be obtained by the same general procedure.

\section{Universality of Shor's Basis}

\subsection{Equivalence between $\{\wedge_1({\sigma_z}^\frac{1}{2}),H\}$
and Shor's basis'}
Kitaev (\cite{kitaevQC}, Lemma 4.6)
has provided a proof for the universality of the
basis $Q_1\equiv\{\wedge_1(\sqrtsz), H\}$.  This basis is equivalent to Shor's
fault-tolerant basis $\{\wedge_2(\sigma_x), \sqrtsz, H\}$, as 
observed in~\cite{kitaevQC}.
We give one demonstration of this equivalence here for
completeness.  This equivalence was also showed independently by Aharonov and
Ben-Or (in journal version of \cite{aharonov1}).

\begin{figure}[h]
\center
\setlength{\unitlength}{0.00083333in}
\begingroup\makeatletter\ifx\SetFigFont\undefined%
\gdef\SetFigFont#1#2#3#4#5{%
  \reset@font\fontsize{#1}{#2pt}%
  \fontfamily{#3}\fontseries{#4}\fontshape{#5}%
  \selectfont}%
\fi\endgroup%
{\renewcommand{\dashlinestretch}{30}
\begin{picture}(3225,1084)(0,-10)
\put(412,1012){\circle*{100}}
\put(2814,612){\circle*{100}}
\put(2212,612){\circle*{100}}
\put(1612,612){\circle*{100}}
\put(1612,1012){\circle*{100}}
\put(2212,1012){\circle*{100}}
\put(2814,1012){\circle*{100}}
\drawline(12,212)(811,212)
\drawline(212,812)(612,812)(612,412)
	(212,412)(212,812)
\drawline(12,612)(212,612)
\drawline(612,612)(812,612)
\drawline(12,1012)(812,1012)
\drawline(411,1013)(411,813)
\drawline(962,662)(1062,662)
\drawline(962,662)(1062,662)
\drawline(962,612)(1062,612)
\drawline(962,612)(1062,612)
\drawline(1412,413)(1812,413)(1812,12)
	(1412,12)(1412,413)
\drawline(2012,413)(2412,413)(2412,12)
	(2012,12)(2012,413)
\drawline(2612,413)(3013,413)(3013,12)
	(2612,12)(2612,413)
\drawline(1212,212)(1412,212)
\drawline(1812,212)(2012,212)
\drawline(2412,212)(2612,212)
\drawline(3013,212)(3212,212)
\drawline(1212,612)(3213,612)
\drawline(1212,1012)(3213,1012)
\drawline(1612,1013)(1612,413)
\drawline(2212,1013)(2212,413)
\drawline(2812,1013)(2812,413)
\put(412,562){\makebox(0,0)[b]{\smash{{{\SetFigFont{8}{9.6}{\familydefault}{\mddefault}{\updefault}${\sigma_z}^\frac{1}{2}$}}}}}
\put(1612,162){\makebox(0,0)[b]{\smash{{{\SetFigFont{8}{9.6}{\familydefault}{\mddefault}{\updefault}$\sigma_z$}}}}}
\put(2212,162){\makebox(0,0)[b]{\smash{{{\SetFigFont{8}{9.6}{\familydefault}{\mddefault}{\updefault}$\sigma_y$}}}}}
\put(2812,162){\makebox(0,0)[b]{\smash{{{\SetFigFont{8}{9.6}{\familydefault}{\mddefault}{\updefault}$\sigma_x$}}}}}
\end{picture}
}
\caption{Constructing $Q_1$ from Shor's basis.}
\label{111111ii}
\end{figure}
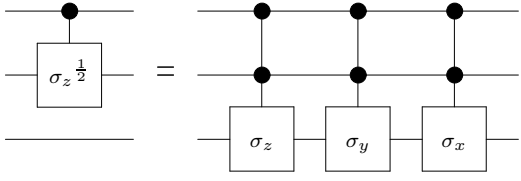

To construct $Q_1$ from Shor's basis, it suffices to construct
$\wedge_1({\sigma_z}^\frac{1}{2})$. First we show that the operations $\wedge_2(\sigma_z)$
and $\wedge_2(\sigma_y)$ can be implemented exactly using gates from Shor's basis
($I_{2^n}$ is the identity operation on $n$\/ qubits.):
\begin{eqnarray*}
H_z &\equiv& H{\sigma_z}^{-\frac{1}{2}}H{\sigma_z}^\frac{1}{2}H \\
\wedge_2(\sigma_z) &=& (I_4\otimes H)\wedge_2(\sigma_x)(I_4\otimes H) \\
\wedge_2(\sigma_y) &=& (I_4\otimes H_z)\wedge_2(\sigma_x)(I_4\otimes H_z),
\end{eqnarray*}
Next, as shown in Figure~\ref{111111ii}, $\wedge_1({\sigma_z}^\frac{1}{2})$ can be
implemented using the identity: $\sigma_x\sigma_y\sigma_z=iI_2$. This reduction also shows
that the universality of $Q_1$ implies the universality of Shor's basis.

\begin{figure}[h]
\center
\setlength{\unitlength}{0.00083333in}
\begingroup\makeatletter\ifx\SetFigFont\undefined%
\gdef\SetFigFont#1#2#3#4#5{%
  \reset@font\fontsize{#1}{#2pt}%
  \fontfamily{#3}\fontseries{#4}\fontshape{#5}%
  \selectfont}%
\fi\endgroup%
{\renewcommand{\dashlinestretch}{30}
\begin{picture}(3225,1083)(0,-10)
\drawline(563,611)(663,611)
\drawline(563,661)(663,661)
\put(1613,1011){\circle*{100}}
\put(2413,1011){\circle*{100}}
\put(2813,1011){\circle*{100}}
\put(1613,611){\circle{200}}
\put(2413,611){\circle{200}}
\put(1213,611){\circle*{100}}
\put(2013,611){\circle*{100}}
\drawline(1012,412)(1412,412)(1412,12)
	(1012,12)(1012,412)
\drawline(1812,412)(2212,412)(2212,12)
	(1812,12)(1812,412)
\drawline(2612,412)(3012,412)(3012,12)
	(2612,12)(2612,412)
\drawline(812,212)(1012,212)
\drawline(1412,212)(1812,212)
\drawline(2212,212)(2612,212)
\drawline(3012,212)(3212,212)
\drawline(813,1011)(3213,1011)
\drawline(813,611)(3213,611)
\drawline(1212,612)(1212,412)
\drawline(1612,1012)(1612,512)
\drawline(2012,612)(2012,412)
\drawline(2412,1012)(2412,512)
\drawline(2812,1012)(2812,412)
\put(1212,162){\makebox(0,0)[b]{\smash{{{\SetFigFont{8}{9.6}{\familydefault}{\mddefault}{\updefault}${\sigma_x}^\frac{1}{2}$}}}}}
\put(2812,162){\makebox(0,0)[b]{\smash{{{\SetFigFont{8}{9.6}{\familydefault}{\mddefault}{\updefault}${\sigma_x}^\frac{1}{2}$}}}}}
\put(2012,162){\makebox(0,0)[b]{\smash{{{\SetFigFont{8}{9.6}{\familydefault}{\mddefault}{\updefault}${\sigma_x}^{-\frac{1}{2}}$}}}}}
\put(212,212){\circle{200}}
\put(213,1011){\circle*{100}}
\put(213,611){\circle*{100}}
\drawline(12,212)(412,212)
\drawline(13,1011)(413,1011)
\drawline(13,611)(413,611)
\drawline(212,1012)(212,112)
\end{picture}
}
\caption{Constructing $\wedge_2(\sigma_x)$ from $Q_1$.}
\label{toffoli}
\end{figure}
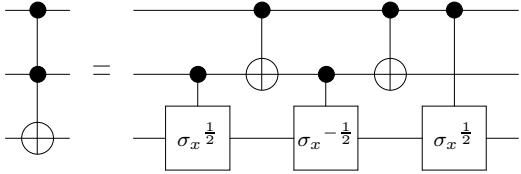

Conversely, to construct Shor's basis from $Q_1$,
it suffices to construct the Toffoli gate $\wedge_2(\sigma_x)$.  Note
that $\wedge_1({\sigma_x}^{\pm\frac{1}{2}})=(I_2\otimes H)
\wedge_1({\sigma_z}^{\pm\frac{1}{2}})(I_2\otimes H)$.  Figure~\ref{toffoli}
gives the circuit construction of Toffoli (see Lemma 6.1
of~\cite{egfqc} for a systematic construction of this circuit).

\subsection{An Alternate Proof}
An alternative proof, which makes use of irrational ``rotations'' about
orthogonal axes, is presented in this section for the universality of
each of the above two basis'.  From either one of these sets, the
following triplet of double-qubit gates is constructible:
\[ G\equiv\{\wedge_1(\sqrtsx),\wedge_1(\sqrtsz),S\} \]
where $S$ is the swap gate: $S|ab\rangle=|ba\rangle$ for any
single-qubit states $|a\rangle$, $|b\rangle$, which can be constructed as
\[ S = \wedge_1(\sigma_x)(H\otimes H)\wedge_1(\sigma_x)
       (H\otimes H)\wedge_1(\sigma_x). \]
For future reference, note that each of the matrices in $G$ are symmetric.
Hence for any matrix $M$ that is constructible from this set, so is its
transpose $M^T$.

It will be shown that any gate in the set
\[
\sum\,\equiv\left(\begin{array}{c|c}
1&\begin{array}{ccc}0&0&0\end{array}\\
\hline
\begin{array}{c}0\\0\\0\end{array}&\mbox{\bf SU}(3)
\end{array}\right)
\]
can be approximated to arbitrary precision by a two-qubit circuit
consisting only of gates from the set $G$.  I.e. the set $G$ under
regular matrix multiplication generates a set dense in $\Sigma$.  From
this set all single-qubit unitary operations {\bf SU}(2) can be
approximated, which along with $\wedge_1(\sigma_x)$, has been
shown \cite{egfqc} to be a universal set of gates.

Though the correspondence is not a strict mathematical correspondence,
it will be useful to make an analogy between real rotations in
3-dimensional space, and gates constructible from $G$.  Define the
following 6 elements of $\left<G\right>$:
\begin{eqnarray*}
\rho_x & \equiv & \wedge_1\!(\sqrtsx)
   \wedge_1\!(\sqrtsz)^{-1} \\
\rho_y & \equiv & S\rho_x^{-1}S \\
\rho_z & \equiv & \wedge_1\!(\sigma_x)
   \rho_y^{-1}\wedge_1\!(\sigma_x) \\
\rho_1 & \equiv & \rho_z^{-1}\wedge_1(\sigma_x)
   \wedge_1(\sqrtsz)\rho_z \\
\rho_2 & \equiv & \rho_x\rho_y \\
\rho_3 & \equiv & \rho_1\rho_2\rho_1^{-1}
\end{eqnarray*}
(Each inverse in the above definitions are obtainable from $G$, since each
element of $G$ is of finite group order.)  Since $\rho_2$ and $\rho_3$
are unitary, they can be unitarily diagonalized:
\begin{eqnarray*}
\rho_2 &=& g_2D(1,1,e^{-i2\pi c},e^{i2\pi c})g_2^{-1} \\
\rho_3 &=& g_3D(1,1,e^{-i2\pi c},e^{i2\pi c})g_3^{-1}
\end{eqnarray*}
where $g_2$, $g_3$ are some unitary matrices (not necessarily
in $\left<G\right>$), $D$ is a diagonal matrix with the given
ordered quadruplet as the entries along the diagonal, and $e^{i2\pi
c}=\frac{1+i\sqrt{15}}{4}$ for some $c\in\mathbb R$.  The minimum monic
polynomial for $e^{i2\pi c}$ over the set of rational numbers is
\[ m_{e^{i2\pi c}}(x) = x^2 - \frac{1}{2}x + 1 \not\in \mathbb Z[x] \]
and thus $c\not\in\mathbb Q$ (see Appendix~\ref{cycloappendix},
Theorem~\ref{cyclotheorem}).  It follows that successive powers of
$\rho_2$ and $\rho_3$ can approximate matrices of the forms
\begin{eqnarray}
\rho_2^{n_2} &\approx&
g_2D(1,1,e^{-i\theta_2},e^{i\theta_2})g_2^{-1}\label{diagtheta2} \\
\rho_3^{n_3} &\approx&
g_3D(1,1,e^{-i\theta_3},e^{i\theta_3})g_3^{-1}\label{diagtheta3}
\end{eqnarray}
for any $\theta_2,\theta_3\in\mathbb R$.  The powers $n_2$ and $n_3$
are functions of $\theta_2$ and $\theta_3$, as well as the desired degree
of accuracy.

The operators $\rho_1$, $\rho_2$, and $\rho_3$ fix the (unnormalized) states
$\left|01\right>-\left|10\right>$, 
$\left|01\right>+\left|10\right>+\left|11\right>$, and
$-\left|01\right>-\left|10\right>+2\left|11\right>$, resp. which form an
orthogonal set of states.  Motivated by considering these 3 operations to
be rotations about 3 orthogonal vectors, a change of basis is performed
into this basis (while mapping the state $\left|00\right>$ to itself).
Under this change of basis, equations (\ref{diagtheta2}) and (\ref{diagtheta3})
are expressed as:
\begin{eqnarray*}
\hat\rho_2^{n_2} &\approx& \left(\begin{array}{cccc}
   1 &  0                      & 0 & 0 \\
   0 &  \cos(\theta_2)         & 0 & \alpha\sin(\theta_2) \\
   0 &  0                      & 1 & 0                       \\
   0 & -\alpha^*\sin(\theta_2) & 0 & \cos(\theta_2)
   \end{array}\right) \\
\hat\rho_3^{n_3} &\approx& \left(\begin{array}{cccc}
   1 & 0                   &  0                     & 0 \\
   0 & \cos(\theta_3)      & -\beta^*\sin(\theta_3) & 0 \\
   0 & \beta\sin(\theta_3) &  \cos(\theta_3)        & 0 \\
   0 & 0                   &  0                     & 1
   \end{array}\right)
\end{eqnarray*}
where $\alpha\equiv\frac{1+2i}{\sqrt5}$ and
$\beta\equiv\frac{1+3i}{\sqrt{10}}$, which like $e^{i2\pi
c}=\frac{1+i\sqrt{15}}{4}$ above, are also seen to not be roots of
unity.

Given any $\gamma\in\mathbb C$, $|\gamma|=1$, define the following
single-parameter group of matrices:
\[ M_\gamma(\theta) \equiv
\left(\begin{array}{rr} \cos(\theta) & -\gamma^*\sin(\theta) \\
\gamma\sin(\theta) & \cos(\theta) \end{array}\right). \]
If $\gamma$ is not a root of unity, then it is straightforward to
show that the set of matrices $\left\{\left.M_\gamma(\theta)^T,
M_\gamma(\theta)\right|\theta\in\mathbb R\right\}$ generates a dense
subset of $\mbox{\bf SU}(2)$.  Given this, and the fact that any element of
$\mbox{\bf SU}(3)$ can be decomposed into a product of $\mbox{\bf SU}(2)$
operations acting
on orthogonal subspaces\cite{tuarg}, it follows that the set
\[ \left\{\hat\rho_2^T,\hat\rho_2,\hat\rho_3^T,\hat\rho_3\right\} \]
generates a dense subset of $\Sigma$.  Since the previous change of
basis bijectively and continuously maps $\Sigma$ onto itself, the
operators $G$ in the original basis generates a dense subset of
$\Sigma$. $\;\;\square$

\section*{Acknowledgments}

We thank A. Kitaev and P. Shor for helpful discussions on the subject
of this paper. We are grateful to D. Aharonov for her comments on an earlier
version of this work.

\appendix

\newtheorem{lemma}{Lemma}[section]

\section{Shor's basis and 
   $\{\, H,\, {\sigma_z}^{\frac{1}{4}},\, \Lambda_1(\sigma_x)\,\}$
   are not equivalent}\label{equiappendix}

In this appendix we show that Shor's basis and our basis
$\{\, H,\, {\sigma_z}^{\frac{1}{4}},\, \Lambda_1(\sigma_x)\,\}$ 
are not equivalent. In fact, every gate in Shor's 
basis can be exactly represented by a circuit over our basis.
First, the following identity shows that our basis can exactly implement 
any gate from the $Q_1$ basis introduced in Section 5.1:
\begin{eqnarray*}
 \Lambda_1({\sigma_z}^{\frac{1}{2}}) 
     & = & \left ( I\otimes{\sigma_z}^{-\frac{1}{4}}\right )\Lambda_1(\sigma_x) \\
     & & \left ( I\otimes{\sigma_z}^{-\frac{1}{4}}\right )\Lambda_1(\sigma_x) \\
     & & \left ( {\sigma_z}^{\frac{1}{4}}\otimes{\sigma_z}^{\frac{1}{2}}\right ).
\end{eqnarray*}
Hence, as proved in the same section, it can exactly implement any gate from
Shor's basis. We prove that the converse is not true. Toward this end, we show that 
the unitary operation ${\sigma_z}^{\frac{1}{4}}$, 
can be computed exactly by our basis but not by
Shor's basis. First we prove a useful Lemma about unitary operations 
computable exactly by Shor's basis. Note that the set of integer complex 
numbers is the set ${\mathbb Z}+i{\mathbb Z}$ of the complex numbers with
integer real and imaginary parts.

\begin{lemma}
Suppose that the unitary operation $U\in\mbox{\bf U}(2^m)$ is the transformation
performed by a circuit $\cal C$ defined over  Shor's basis with $m$ inputs. Then $U$ is 
of the form $\frac{1}{\sqrt{2}^\ell}M$, where $M$ is a $2^m\times 2^m$
matrix with only complex integer entries.
\label{shor-basis-lemma}
\end{lemma}

{\bf Proof.} Suppose that $g_1,\ldots,g_t$ are the gates of $\cal C$.
Each gate $g_j$ can be considered as a unitary operation in 
$\mbox{\bf U}(2^m)$ by acting as an identity operator on the qubits that are not inputs
of $g_j$. Let the matrix $M_j\in\mbox{\bf U}(2^m)$ represent $g_j$. Then
$U=M_t\cdots M_1$. If $g_j$ is a ${\sigma_z}^{\frac{1}{2}}$ gate then $M_j$ is
a diagonal matrix with $1$ or $i$ on its diagonal. If $g_j$ is a Toffoli gate
then $M_j$ is a permutation matrix (which is a 0--1 matrix). Finally, if $g_j$ is a
Hadamard gate, then $M_j=\frac{1}{\sqrt{2}}{M_j}'$, where the entries of
${M_j}'$ are integers.
This completes the proof. $\square$

Now since 
${\sigma_z}^{\frac{1}{4}}=\frac{1}{\sqrt{2}}\begin{pmatrix}
   \sqrt{2} & 0 \\ 0 & 1+i \end{pmatrix}$, by Lemma \ref{shor-basis-lemma}
it cannot be realized exactly by gates from Shor's basis.

\newtheorem{theorem}{Theorem}[section]

\section{The Cyclotomic/Rational Number Theorem}\label{cycloappendix}
\begin{theorem}\label{cyclotheorem}
For any $c\in\mathbb R$, the following two statements are logically
equivalent:

\begin{itemize}
\item The minimum monic polynomial $m_\alpha(x)\in\mathbb Q[x]$ for
$\alpha\equiv e^{i2\pi c}$ exists and is cyclotomic.
\item $c\in\mathbb Q$.
\end{itemize}

\end{theorem}
{\it Proof:}\\
A number of algebraic theorems will be taken for granted in this proof,
in particular, properties of cyclotomic polynomials $\Phi_n(x)$.  See,
for instance, Dummit and Foote\cite{aa} for a more thorough discussion
of these polynomials, as well as general properties of polynomial rings.

Assume $m_\alpha(x)$ exists and $m_\alpha(x)=\Phi_n(x)$ for some
$n\in\mathbb{Z}^+$.
\begin{eqnarray*}
0 &=\footnotemark[1]& m_\alpha(\alpha) \\
  &=\footnotemark[2]& \Phi_n(\alpha) \\
  &=\footnotemark[3]& \prod_{d|n}\Phi_d(\alpha) \\
  &=\footnotemark[4]& \alpha^n-1 \\
  &=\footnotemark[5]& e^{i2\pi cn}-1.
\end{eqnarray*}
$nc\in\mathbb Z$.  Thus $c\in\mathbb Q$.
\footnotetext[1]{Definition of $m_\alpha(x)$}
\footnotetext[2]{By assumption.}
\footnotetext[3]{0 times anything is 0.}
\footnotetext[4]{Property of cyclotomic polynomials.}
\footnotetext[5]{Definition of $\alpha$.}

Conversely, assume $c\in\mathbb{Q}$.
$c=\frac{p}{q}$ for some $p,q\in\mathbb Z$.
$m_\alpha(x)$ exists, since $\alpha^q-1=e^{i2\pi cq}-1=e^{i2\pi p}-1=0$.
Moreover, $m_\alpha(x)$ divides $x^q-1=\prod_{d|q}\Phi_d(x)$ in
$\mathbb Q[x]$.
$m_\alpha(x)\propto\Phi_n(x)$ for some $n|q$.
Since both are monic, $m_\alpha(x)=\Phi_n(x).\;\;\square$

\end{document}